\documentstyle[epsf]{mn}
\def\lsim{\mathrel{\mathpalette\versim<}}

\def\versim#1#2{\vcenter{\offinterlineskip
	\ialign{$#1\hfil##\hfil$\crcr#2\crcr\sim\crcr } }}

\newcommand{\x}{\mbox{$\times$}}
\newcommand{\e}[1]{\x10^{#1}}

\newcommand{\Mo}{\mbox{$M_\odot$}}
\newcommand{\Ro}{\mbox{$R_\odot$}}

\newcommand{\ns}{\mbox{$n_\star$}}
\newcommand{\Mbh}{\mbox{$M_{\rm bh}$}}
\newcommand{\Mtot}{\mbox{$M_{\rm sys}$}}

\newcommand{\Mw}{\mbox{$M_{\rm w}$}}
\newcommand{\Rw}{\mbox{$R_{\rm w}$}}
\newcommand{\Rm}{\mbox{$R_{\rm mass}$}}
\newcommand{\Rt}{\mbox{$R_{\rm tidal}$}}
\newcommand{\dMw}{\mbox{$\dot{M}_{\rm w}$}}
\newcommand{\Rc}{\mbox{$R_{\rm \/C}$}}
\newcommand{\Nc}{\mbox{$N_{\rm \/C}$}}

\newcommand{\Rb}{\mbox{$R_\star$}}
\newcommand{\vb}{\mbox{$v_\star$}}

\newcommand{\Mb}{\mbox{$M_\star$}}
\newcommand{\MBS}{\mbox{$M_{\rm BS}$}}
\newcommand{\Lion}{\mbox{$L_{\rm ion}$}}
\newcommand{\EBV}{\mbox{$E_{\rm \/B-V}$}}
\newcommand{\rin}{\mbox{$r_{\rm in}$}}
\newcommand{\rout}{\mbox{$r_{\rm out}$}}
\newcommand{\fBS}{\mbox{$f_{\rm \/BS}$}}

\newcommand{\LtoM}{\mbox{$L_{\rm ion}/\Mbh$}}
\def  \La          {\ifmmode {\rm Ly}\alpha \else Ly$\alpha$\fi}
\def  \Lalpha      {\ifmmode {\rm Ly}\alpha\,\lambda1215
		    \else Ly$\alpha$\,$\lambda1215$\fi}
\def  \Ka          {\ifmmode {\rm K}\alpha \else K$\alpha$\fi}
\def  \Lb          {\ifmmode {\rm L}\beta \else L$\beta$\fi}
\def  \Ha          {\ifmmode {\rm H}\alpha \else H$\alpha$\fi}
\def  \Halpha      {\ifmmode {\rm H}\alpha\,\lambda6563
		    \else H$\alpha$\,$\lambda6563$\fi}
\def  \Hb          {\ifmmode {\rm H}\beta \else H$\beta$\fi}
\def  \Hbeta       {\ifmmode {\rm H}\beta\,\lambda4861
		    \else H$\beta$\,$\lambda4861$\fi}
\def  \Pa          {\ifmmode {\rm P}\alpha \else P$\alpha$\fi}
\def  \HeI         {\ifmmode {\rm He}\,{\sc i}\,\lambda5876
		    \else He\,{\sc i}\,$\lambda5876$\fi}
\def  \HeII        {\ifmmode {\rm He}\,{\sc ii}\,\lambda1640
		    \else He\,{\sc ii}\,$\lambda1640$\fi}
\def  \CIII        {\ifmmode {\rm C}\,{\sc iii}\,\lambda977
		    \else C\,{\sc iii}\,$\lambda977$\fi}
\def  \CIIIb       {\ifmmode {\rm C}\,{\sc iii]}\,\lambda1909
		    \else C\,{\sc iii]}\,$\lambda1909$\fi}
\def  \CIV         {\ifmmode {\rm C}\,{\sc iv}\,\lambda1549
		    \else C\,{\sc iv}\,$\lambda1549$\fi}
\def  \bOIIIbA     {\ifmmode {\rm [O}\,{\sc iii]}\,\lambda4363
		    \else [O\,{\sc iii]}\,$\lambda4363$\fi}
\def  \bOIIIbB     {\ifmmode {\rm [O}\,{\sc iii]}\,\lambda5007
		    \else [O\,{\sc iii]}\,$\lambda5007$\fi}
\def  \OIIIb       {\ifmmode {\rm O}\,{\sc iii]}\,\lambda1663
		    \else O\,{\sc iii]}\,$\lambda1663$\fi}
\def  \OVb         {\ifmmode {\rm O}\,{\sc v]}\,\lambda1218
		    \else O\,{\sc v]}\,$\lambda1218$\fi}
\def  \OVI         {\ifmmode {\rm O}\,{\sc vi}\,\lambda1035
		    \else O\,{\sc vi}\,$\lambda1035$\fi}
\def  \OIVb        {\ifmmode {\rm O}\,{\sc iv]}\,\lambda1402
		    \else O\,{\sc iv]}\,$\lambda1402$\fi}
\def  \bOIIb       {\ifmmode {\rm [O}\,{\sc ii]}\,\lambda3727
		    \else [O\,{\sc ii]}\,$\lambda3727$\fi}
\def  \bOIb        {\ifmmode {\rm [O}\,{\sc i]}\,\lambda6300
		    \else [O\,{\sc i]}\,$\lambda6300$\fi}
\def  \NV          {\ifmmode {\rm N}\,{\sc v}\,\lambda1240
		    \else N\,{\sc v}\,$\lambda1240$\fi}
\def  \NIVb        {\ifmmode {\rm N}\,{\sc iv]}\,\lambda1486
		    \else N\,{\sc iv]}\,$\lambda1486$\fi}
\def  \NIIIb       {\ifmmode {\rm N}\,{\sc iii]}\,\lambda1750
		    \else N\,{\sc iii]}\,$\lambda1750$\fi}
\def  \MgII        {\ifmmode {\rm Mg}\,{\sc ii}\,\lambda2798
		    \else Mg\,{\sc ii}\,$\lambda2798$\fi}
\def  \bNeVb       {\ifmmode {\rm [Ne}\,{\sc v]}\,\lambda3426
		    \else [Ne\,{\sc v]}\,$\lambda3426$\fi}
\def  \NeVIII      {\ifmmode {\rm Ne}\,{\sc viii}\,\lambda774
		    \else Ne\,{\sc viii}\,$\lambda774$\fi}
\def  \SiIV        {\ifmmode {\rm Si}\,{\sc iv}\,\lambda1397
		    \else Si\,{\sc iv}\,$\lambda1397$\fi}
\def  \bFeXb       {\ifmmode {\rm [Fe}\,{\sc x]}\,\lambda6734
		    \else [Fe\,{\sc x]}\,$\lambda6734$\fi}
\def  \bFeXIb      {\ifmmode {\rm [Fe}\,{\sc xi]}\,\lambda7892
		    \else [Fe\,{\sc xi]}\,$\lambda7892$\fi}
\def  \FeII        {\ifmmode {\rm Fe}\,{\sc ii}\,
		    \else Fe\,{\sc ii}\,\fi}
\def  \MgIIw       {\ifmmode {\rm Mg}\,{\sc ii}\,\lambda2798{\rm+}\FeII
		    \else Mg\,{\sc ii}\,$\lambda2798$+$\FeII$\fi}
\def  \MgIIc       {\ifmmode {\rm Mg}\,{\sc ii}\,\lambda2798{\rm-}\FeII
		    \else Mg\,{\sc ii}\,$\lambda2798$-$\FeII$\fi}

\load{\normalsize}{\sc}

\begin{document}

\title[Bloated Stars as AGN Broad Line Clouds]
      {Bloated Stars as AGN Broad Line Clouds:\\
      The Emission Line Profiles}
\author[T. Alexander and H. Netzer]
      {Tal Alexander\thanks{E-mail address: tal@wise.tau.ac.il} and 
       Hagai Netzer\thanks{E-mail address: netzer@wise.tau.ac.il}\\
       School of Physics and Astronomy and the Wise observatory\thanks{
       Of the Beverly and Raymond Sackler Faculty of Exact Sciences.}, 
       Tel-Aviv University, 
       Tel-Aviv 69978, Israel.}
\maketitle 
\begin{abstract} 
The `Bloated Stars Scenario' proposes that AGN broad line emission
originates in the winds or envelopes of bloated stars (BS). Alexander and
Netzer (1994) established that $\sim 5\e{4}$ BSs with dense, decelerating
winds can reproduce the observed emission line spectrum and avoid rapid
collisional destruction. Here, we use the observed properties of AGN line
profiles to further constrain the model parameters.  In the BS model, the
origin of the broad profiles is the stellar velocity field in the vicinity
of the central black hole. We use a detailed photoionization code and a
model of the stellar distribution function to calculate the BS emission
line profiles and compare them to a large sample of AGN $\CIV$, $\CIIIb$
and $\MgII$ profiles. We find that the BSs can reproduce the general shape
and width of typical AGN profiles as well as the line ratios if (i) The
ionizing luminosity to black hole mass ratio is low enough. (ii) The broad
line region size is limited by some cutoff mechanism. (iii) The fraction of
the BSs in the stellar population falls off roughly as $r^{-2}$. (iv) The
wind density and/or velocity are correlated with the black hole mass and
ionizing luminosity. Under these conditions the strong tidal forces near
the black hole play an important role in determining the line emission
properties of the BSs. Some discrepancies remain: the calculated BS
profiles tend to have weaker wings than the observed ones, and the
differences between the profiles of different lines are somewhat smaller
than those observed.
\end{abstract}

\section{Introduction}
  
    The observed properties of active galactic nuclei (AGN) lead to the
conclusion that the broad line emission originates in numerous small, cold
and dense gas concentrations. These objects are labeled ``clouds'',
although their true nature remains unknown. A basic challenge for any broad
line region (BLR) model is to specify the physical mechanism that protects
the clouds from rapid disintegration in the AGN's extreme environment, or
else specify a source for their continued replenishment. The bloated stars
(BSs) model \cite{Edwards,Mathews,SN,Penston,Kazanas} proposes that the
lines are emitted from the winds or mass loss envelopes of giant stars. The
star provides both the gravitational confinement and the mass reservoir for
replacing the gas that evaporates from the envelope to the interstellar
medium (ISM). The BS model is further motivated by the lack of
observational evidence for net radial motion in the BLR (e.g. Maoz et
al. 1991, Wanders et al. 1995). This is consistent with virialized motion
in the gravitational potential of the nucleus.

   The BS model was recently studied by Alexander \& Netzer (1994, paper
I), which used a detailed photoionization code to calculate the emission
line spectrum of various simple BS wind models. The integrated emission
line spectrum was calculated by combining the line emissivity of the BSs
with theoretical models of the stellar distribution function (DF). The
properties of the models were compared to the mean AGN line ratios and to
estimates of the BLR size and line reddening. The mean observed $\La$
equivalent width was used to determine the number of BSs and their fraction
in the stellar population and to estimate the collisional mass loss rate
and its effect on the ISM electron scattering optical depth.

   The main result of paper I is that there are BS wind models that can
reproduce the emission line ratios to a fair degree with only a small
fraction of BS in the stellar population. However, these models do not
resemble winds of normal supergiants. The photoionization calculations show
that the emission line spectrum is dominated by the conditions at the outer
boundary of the line emitting zone of the wind. The successful BS models
are those with dense envelopes that have small density gradients. In this
case the wind boundary is set by tidal forces near the black hole and by
the finite mass of the wind at larger radii. Only $\sim 5\x10^4$ such BSs
($< 1\%$ of the BLR stellar population) are required for reproducing the
BLR emission. As a result, the collision rate is reasonably small ($< 1
\Mo$/yr). On the other hand, lower-density models or those with a steep
density gradients are ruled out because they emit strong broad forbidden
lines, which are not observed in AGN. In addition, their low line
emissivity requires orders of magnitude more BSs in the BLR. This leads to
rapid collisional destruction, very high mass loss rate and very short BS
lifetimes.  Even if the BSs are continuously created, the ISM electron
scattering optical depth is very large, contrary to what is observed.

   In this second part of the work, we combine our line emission
calculations with the dynamics of the stellar cluster to obtain the
predicted line profiles for the BS model. This provides a second, powerful
test that can, when compounded with observed line profiles, put more
constraints on this idea. Section \ref{s:model} describes the components of
the model: the BSs, the ionizing continuum, the stellar distribution
function and the black hole mass. Section \ref{s:prof} describes the line
profile calculations and presents a large sample of observed profiles. In
section \ref{s:result} we compare the profiles of various BS models to the
observed ones. A discussion and a summary of our conclusions are presented
in section~\ref{s:discuss}.


\section{The model}
\label{s:model}

   The observed properties of AGN profiles broadly fall into the following
categories:
\begin{enumerate}
\item The line width distribution in the AGN population.
\item The profile shapes.
\item The differences between profiles of different lines.
\item Profile asymmetries.
\item Profile shifts relative to the systemic redshift.
\end{enumerate}
A successful BLR model should be able to explain all these properties. The
BS model, in its current level of detail, does not attempt to explain the
profile asymmetries nor the relative line shifts. Nevertheless, even the
remaining profile properties set strong constraints on the BS model. We
proceed on the premise that the asymmetries and relative shifts can be
eventually explained either within the framework of the BS model or by the
existence of another line emitting component in the BLR.

   The integrated line emission is a spatial average of the line emissivity
in the BLR volume, weighted by the stellar density.  Consequently, the line
ratios do not impose stringent constraints on the spatial distribution of
the gas. The line profile, $L_u$ ($u$ is the line of sight projection of
the 3D velocity $\bmath{v}$), is a finer probe of the BLR structure, since
unlike the total emission, it is integrated only over gas which has line of
sight velocity $u$.

   The motion of the BSs follows the velocity field of the stellar system
that surrounds the black hole.  The input parameters that are required for
calculating the line profiles are the structure of the BS wind, the black
hole mass, the stellar DF and the distribution of the BSs in the normal
stellar population. A detailed description of these ingredients can be
found in paper I. Here we summarize the relevant properties and list the
new features that were introduced into the model.

\subsection{The bloated stars}
\label{s:BS1}
 
   There are to date no reliable models for the structure of massive winds
in the presence of an intense external radiation field (see however the
recent work by Scoville \& Norman \shortcite{SN2} and Hartquist et al
\shortcite{HDDRWW}). At this stage, our purpose is to understand the
conditions that are required for the BSs to be consistent with both the
observed BLR properties and the properties of the stellar population. We
therefore model the BS by a simple structure with no claim to
hydrodynamical self-consistency and without specifying the process that
drives the wind. The validity of this approximation was demonstrated in
paper I by showing that the line emission does not depend on the wind
internal structure, but predominantly on the conditions at the boundary of
its line emitting zone. The BS is modeled by two components: a giant star
of radius $\Rb = 10^{13}$ cm and a mass of $\Mb = 0.8\Mo$, emitting a
spherically symmetric wind of radius $\Rw$ which contains an additional
mass of up to $\Mw = 0.2\Mo$. The wind density and velocity are linked by
the continuity equation
\begin{equation} 
\label{e:cont}
\dMw = 4\pi R^2 v(R) N(R)\,.
\end{equation}
where $\dMw = 10^{-6} \Mo$/yr is the mass loss rate and $N$ is the hydrogen
number density. The wind structure is parameterized by a power-law velocity
field:
\begin{equation} 
\label{e:plawv}
v(R) = \vb (R/\Rb)^{-\alpha}\,,
\end{equation}
where $\vb$ is the velocity at the surface of the star. This family of
models spans a two-parameter space, $(\alpha,\vb)$, which was investigated
extensively in paper I. Here we will concentrate on the values that best
reproduce the observed emission line spectrum, $\alpha = 1/2$ (free
falling flow) and $\vb \sim 10^4$ cm/s (slow and dense wind, $10^8 \lsim N
\lsim 10^{12}$ cm$^{-3}$).
    
   The size of the wind, and hence its boundary density is determined by
three competing physical processes. Tidal disruption of the outer layers of
the wind by the black hole limits the size of the wind to
\begin{equation}
\label{e:rt}
\Rt(r)= X_{\rm tidal} (M_\star/\Mbh)^{1/3}r\,,
\end{equation}
where $\Mbh$ is the black hole mass, $r$ is the distance from the black
hole and $X_{\rm tidal}$ is a factor of order unity (we assume $X_{\rm
tidal} = 2$). We will use the term ``tidally-limited'' to designate such
BSs.  The finite mass in the wind sets an upper limit on the wind size,
$\Rm$,
\begin{equation} 
\label{e:rmass}
\int_{\Rb}^{\Rm} \frac{\dMw}{v}dR = \Mw\,.
\end{equation}
We will use the term ``mass-limited'' to designate such BSs. The third
process is Comptonization \cite{Kazanas}, whereby the central ionizing
continuum heats the outer layers of the wind, ionizes them completely and
reduces their optical depth to zero. The radiation penetrates into the
denser parts of the wind until the density rises above a critical value
$\Nc$, and an ionization equilibrium is established at the Comptonization
radius $\Rc$. The emission line spectrum is determined by the ionization
parameter, $U$, which is essentially the ratio of the ionizing flux and gas
density at the gas surface. The results of paper I show that the gas at the
Comptonized wind boundary has a high $U$. This results in strong emission
of unobserved forbidden lines, and therefore this process must be
suppressed. The high gas density and small density gradient of the slow and
dense wind models achieve this suppression by having $\Rc > \Rt$ or $\Rc >
\Rm$.

\subsection{The galactic nucleus}

   The stellar distribution function that we use is based on the numeric
results of Murphy, Cohn \& Durisen \shortcite{MCD} (henceforth MCD), which
follow the evolution of a multi-mass coeval stellar cluster in the presence
of a central black hole. The black hole mass grows as it accretes mass from
the stars, both directly, by tidal disruption, and indirectly, from mass
loss in the course of stellar evolution and of inelastic stellar
collisions. The MCD calculations assume spherical symmetry and an initial
Plummer DF with a seed black hole of $10^4 \Mo$.

   In this work we use MCD model 2B, which has an initial central density
of $7\x10^7\Mo$/pc$^3$, $\sim 3\x10^8\Mo$ within the inner 1 pc and a total
mass of $\Mtot = 8.5\x10^8\Mo$ within 100 pc. The luminosity is due to
spherical accretion with rest-mass to luminosity conversion efficiency of
0.1. The luminosity reaches a peak of $2.4\x10^{46}$ erg/s at $4\x10^8$
yr. The black hole reaches a mass of $5.5\x10^8\Mo$ after 15 Gyr. We study
this system at two epochs. The first is at $t_0 = 3\x10^8$ yr, when $\Mbh =
8\x10^7\Mo$, $L_{\rm ion} = 7\x10^{45}$ erg/s and the luminosity is
Eddington-limited. This young system, which was the one studied in paper I,
has the luminosity of a bright quasar. The second is at $t_0 = 10^9$ yr,
when $\Mbh = 1.9\x10^8\Mo$, $L_{\rm ion} = 3.6\x10^{44}$, the luminosity is
sub-Eddington and determined by the mass loss rate from the stellar
system. This evolved system has the luminosity of a luminous Seyfert 1.

   MCD assume, for simplicity, that the velocity DF can be approximated
by the Maxwell-Boltzmann DF
\begin{equation} 
\label{e:MB}
{\rm DF}(r,v,t) = 
   \left(\frac{3}{2\pi}\right)^{3/2} \frac{\ns(r,t)}{v_0(r,t)^3} 
   \exp\left[-\frac{3}{2}\left(\frac{v}{v_0(r,t)}\right)^2\right]\,,
\end{equation} 
where $n_\star$ is the stellar density and $v_0$ the r.m.s velocity. Thus,
the velocity field is fully specified by $v_0(r,t)$, which MCD calculate
numerically by following the time evolution of the system.

   Our calculations use an approximate $\ns(r,t)$\footnote{Note that here,
as opposed to paper I, we use $\ns$ to designate the total stellar
population, BSs included.}, which is based on the density function of the
$0.8\Mo$ mass bin stars (that being the mass bin closest to $1\Mo$ in MCD
figure 6, see paper I). The density function $n_{0.8}(r)$ is given in MCD
for several values of $t_0$. We interpolate it to any $t_0$ and then
approximate $\ns$ by
\begin{equation}  
\ns(r,t_0) = n_{0.8}(r,t_0) 
             \frac{\Mtot-\Mbh}{\Mo4\pi\int r^{\prime2}n_{0.8}dr^\prime}\,,
\end{equation} 
that is, by approximating that the ratio between the total stellar density
and that of the $0.8\Mo$ mass bin stars, ($\sim 4$), does not depend on $r$
and that the mean stellar mass is $1\Mo$.  For $v_0(r,t)$ we use a
semi-analytic approximation, which is derived from the following
arguments. We model the stellar core at times $t_0 = 3\x10^8$ and $10^9$
yr, which are both earlier than the collisional time-scale of the initial
Plummer distribution, $t_{\rm c} = 1.2$ Gyr. The relevant dynamical length
scale in the nucleus is the black hole dynamical radius, $r_{\rm dyn} =
G\Mbh/v^2_{\rm init}$, where $v_{\rm init}$ is the initial Plummer r.m.s
velocity ($v_{\rm init}= 1360$ km/s for MCD model 2B). At $r \ll r_{\rm
dyn}$, the initial DF evolves rapidly due to the frequent stellar
collisions in the deep potential well, so that by $t_0$ the velocity field
is dominated by the black hole. Consequently, only the black hole mass,
rather than the detailed form of $n_\star(r,t)$, is required for
calculating $v_0$. The spatial redistribution of the stellar mass at $r \ll
r_{\rm dyn}$, whether by the evolution of the DF or by the growth of the
black hole, does not affect the potential at $r > r_{\rm dyn}$ because the
mass distribution is spherically symmetric. Consequently, if $t_0 < t_{\rm
c}$, the initial Plummer DF is almost unchanged except for the small
evolutionary mass loss. The MCD models assume that the initial mass
function is independent of $r$, and therefore the evolutionary mass loss is
reflected only in a small, spatially homogeneous decrease of the initial
stellar density.

   At small $r$, the potential of the black hole dominates and $v_0^2 =
\gamma G\Mbh/r$, where $\gamma$ is of order unity. The width of the line
profiles strongly depends on $\gamma$ and it is therefore important to
estimate it as reliably as possible. Cohn \& Kulsrud \shortcite{CK} find
that $\gamma \sim 1.2$ at $0.001 < r/r_{\rm dyn} < 0.1$. Bahcall \& Wolf
\shortcite{BW} similarly find that $\gamma \ga 12/11$ at $r < r_{\rm dyn}$
and that it increases by up to a factor of 2 as $r$ decreases. David et al
\shortcite{DDC} and MCD (B. Murphy, private communication) find that
$\gamma$ approaches 2 as $r$ decreases. This result can be explained
qualitatively in terms of the orbits of the stars near the black hole. If
these stars were in bound circular Keplerian orbits, $\gamma$ would have
been 1. However, such bound stars are quickly destroyed by stellar
collisions and tidal disruption. The only stars that can exist at such
small radii are those that spend there only a very short fraction of their
period and have high enough angular momenta to avoid falling into the black
hole. Such orbits are marginally bound, nearly parabolic Keplerian orbits,
which have $\gamma = 2$. At larger radii the stellar destruction proceeds
more slowly, the fraction of bound stars increases and therefore $\gamma$
decreases.

   The r.m.s. velocity of the initial Plummer distribution is
\begin{equation}
v_0 = \sqrt{\frac{\Psi(r)}{2}}\,, 
\end{equation} 
where $\Psi$ is the negative of the gravitational potential
\begin{equation}
\Psi = \frac{G \Mtot}{\sqrt{r_0^2+r^2}}\,,
\end{equation}
and $r_0 = 1$ pc is the core radius \cite{BT}. We propose to approximate
the r.m.s velocity by the simplest expression that approaches the correct
limits at small and large $r$,
\begin{equation}
\label{e:v0}
v_0^2(r) = \frac{G(\Mtot-\Mbh)}{2\sqrt{r_0^2+r^2}}
      + \gamma(r)\frac{ G \Mbh}{r}\,,
\end{equation} 
where $\gamma(r)$ is a smooth function that approaches 2 at small radii and
1 at large radii.

   We approximate $\gamma(r)$ by assuming that dynamically, the stellar
population near the black hole is composed of two types of stars. A
fraction $f_{\rm b}$ of the stars are bound to the black hole in circular
orbits and have $\gamma_b = 1$, while the remaining $f_u = 1-f_{\rm b}$ of the
stars are unbound, in parabolic orbits with $\gamma_u = 2$. We further
assume that the velocity distributions of the bound and unbound stars are
Maxwellian with $v_{0,b}^2$ and $v_{0,u}^2$ of the same form as
equation~\ref{e:v0}, with $\gamma_b $ and $\gamma_u$, respectively.  It
then follows that the total population has $v_0^2 = f_b v_{0,b}^2+f_u
v_{0,u}^2$ and that
\begin{equation}
\label{e:gr}
\gamma(r) = f_b\gamma_b+f_u\gamma_u = 2-f_b\,.  
\end{equation} 
We assume that initially all the stars are bound. With time, inelastic
collisions and tidal disruption by the black hole deplete the fraction of
bound stars. The MCD results show that inelastic stellar collisions is the
dominant mass loss process for the high density stellar cores which are
studied here. We therefore neglect the tidal disruptions as a mass loss
process. 

   $f_b(t_0)$ is equal to the probability of a bound star to survive up to
time $t_0$. At each inelastic encounter, which occur at a rate $q_c$, the
star loses on average a fraction $\kappa$ of its mass. By approximating
that the fractional collisional mass loss rate, $\kappa q_c$, is constant
in time and equal to its value at $t_0$, we obtain
\begin{equation} 
f_b = \exp(-\kappa q_{\rm c} t_0)\,.
\end{equation} 
The collisional rate for equal mass and radius stars with a Maxwellian DF
is \cite{BT}
\begin{equation}
\label{e:qcoll} 
q_{\rm c}(t_0) = 16 \sqrt{\frac{\pi}{3}} \ns v_0 R_\star^2 
                    \left(1+\frac{3 v_{\rm esc}^2}{4v_0^2}\right)\,,
\end{equation} 
where $R_\star$ is the stellar radius, $v_{\rm esc}$ is the escape velocity
from the star and the second term in the brackets is the Safronov number,
which takes into account the enhancement of the collision rate by
gravitational focusing. For $\kappa$ we use the approximate form suggested
by MCD,
\begin{equation} 
\label{e:kappa}
\makebox[1.0 in][l]{$\kappa = \kappa_\infty 10^{-v_{\rm esc}/v_0}\,,$}
\makebox[1.0 in][r]{$\kappa_\infty= 0.1\,.$}
\end{equation}
Both $q_{\rm c}$ and $\kappa$ are functions of $\gamma$ (through $v_0$) and
therefore $\gamma(r)$ (equation~\ref{e:gr}) is the root of the equation
\begin{equation} 
\gamma = 2-\exp(-\kappa(\gamma) q_{\rm c}(\gamma) t_0)\,,
\end{equation}	
which is solved for $\gamma$ numerically.

\begin{figure} 
   \centering \epsfxsize=240pt
   \epsfbox{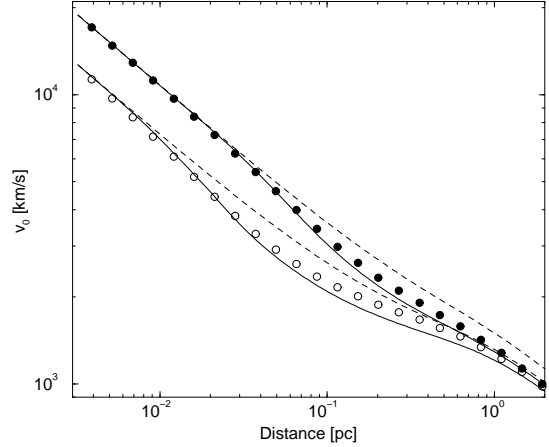} 
   \caption{The analytic approximation of the  r.m.s stellar velocity,
            $v_0$, compared to the MCD results. 
            Solid line: The approximate $v_0$ (equation~\protect\ref{e:v0})
            of the main sequence stars.
            Dashed line: The approximate $v_0$ of the marginally bound
            ($\gamma = 2$) stars only. 
            Open circles: $v_0$ calculated by MCD for model 2B at 
            $t_0 = 6.3\x10^7$ yr, when $\Mbh =6.0\x10^7\Mo$. 
            Full circles: $v_0$ calculated by MCD for model 2B at 
            $t_0 = 5.3\x10^8$ yr, when $\Mbh = 1.3\x10^8\Mo$. 
            (Courtesy of B. Murphy)}
\label{f:v0}
\end{figure}

   Figure~\ref{f:v0} compares the approximated $v_0$ to that calculated by
MCD for model 2B\footnote{The numeric results displayed in Fig.~\ref{f:v0}
were calculated for an initial mass function in the range $0.12$ to $12\Mo$
instead of the original range of $0.3$ to $30\Mo$ in MCD.}  at two epochs
(B. Murphy, private communication). The approximation tends to
under-estimate the velocity at larger $r$ and slightly over-estimate it at
small $r$. These trends appear also for the younger or less massive MCD
models. However, in all these cases the fractional error in the
approximation is less than 10\%, which we consider to be well within the
joint uncertainties of the BS model and the MCD results.

   Finally, we note that BSs are expected to undergo more collisions than
the smaller main sequence (MS) stars. It is therefore possible that the BSs
will not share the same velocity field as the general stellar population
but are all marginally bound with $\gamma = 2$. Figure~\ref{f:v0} shows
that this may result in an increase of up to $\sim 20\%$ in the BS r.m.s
velocity in young systems when $t_0$ is much smaller than the MS
collisional time-scale, but less so in more evolved systems.

\subsection{New features of the BS model}

\subsubsection{The fraction of BSs}

   In paper I, we assumed that the fraction of the BS in the stellar
population, $\fBS$, is constant and independent of $r$. This is
obviously only a crude approximation. BSs are not observed in normal
stellar environments, such as are, presumably, the host galaxies of AGN.
This implies that $\fBS$ must approach zero at large $r$.  The results
of section~\ref{s:result} below indicate that the form $\fBS \propto
r^{-2}$ may be necessary for reproducing the shape of the observed
profiles. We therefore model here also the case of
\begin{equation}
\label{e:fBSr2}
\fBS(r) = \min(f_{\max},f_{\max}\left(\frac{r_f}{r}\right)^2)\,,
\end{equation} 
where $f_{\max} \le 1$ is an upper bound on the fraction of BSs in the
stellar population and $r_f$ is the distance where the BS fraction
reaches this bound. $f_{\max}$ is a free parameter of the BS model. We
will make the arbitrary assumption that $f_{\max} \sim 1/4$, the
fraction of the $0.8\Mo$ mass bin stars in the stellar population.
For the model discussed below, $v_0(r_f) > 10000$ km s$^{-1}$. Since
BSs with such a high velocity contribute very little to the line
emission, for all practical purposes $\fBS \propto r^{-2}$.

\subsubsection{Collisional mass loss}

   Collisional mass loss sets a strong constraint on BS wind
models. Reliable estimates of the mass loss rate are especially important
for BS distributions that are concentrated towards the black hole.  Here we
refine the simple estimates used in paper I for the collisional mass loss
rate per volume by using equations~\ref{e:qcoll} and \ref{e:kappa} and by
taking into account the exact amount of mass in the wind, rather than the
typical $\Mw = 0.2\Mo$. Thus, for BS--BS collisions
\begin{eqnarray}
\dot{\rho}_{\rm BS\,coll}(r) & = & \fBS(r)\ns(r) [
   q_{{\rm c},\star}\kappa_\star\Mb+ \\ \nonumber
&& (q_{\rm c,w}-q_{{\rm c},\star})\kappa_{\rm w}\Mw(\Rw)]\,,
   \nonumber
\end{eqnarray}
where the collisions of the giant star and of the wind are treated
separately, $q_{\rm c,\star}$ and $\kappa_\star$ are calculated with
$\Mb$, $\Rb$ and $q_{\rm c,w}$ and $\kappa_{\rm w}$ are calculated with
$\Mb+\Mw(\Rw)$, $\Rw$. Similarly, for the MS--MS collisions
\begin{equation} 
\dot{\rho}_{\rm MS\,coll}(r) = [1-\fBS(r)]\ns(r)q_{\rm c}\kappa\Mo\,,
\end{equation} 
where $q_{\rm c}$ and $\kappa$ are calculated with 1 $\Mo$ and 1 $\Ro$.

\subsubsection{Reddening correction}

   The amount of intrinsic reddening in AGN spectra is still an
unresolved issue \cite{MacAlpine,BNW}. We assume the ``screen hypothesis'',
whereby the dust lies just outside the line emitting region and
intercepts all the observed spectrum, continuum as well as lines. In
paper I we estimated the extinction coefficient, $\EBV$, from the
difference between the calculated and the mean observed $\La/\Hb$ ratio.
We then corrected all the other lines accordingly. This procedure does
not allow for the fact that there is actually a range of observed
$\La/\Hb$ ratios. It is also vulnerable to possible inaccuracies in the
$\La$ and $\Hb$ radiative transfer method that is used by the
photoionization code. This was not a serious problem in paper I, since
the models that succeeded in reproducing the emission line spectrum
required only a small reddening correction, while the other models were
ruled out by excessive collisional mass-loss. Here we relax the
reddening criterion and determine $\EBV$ within the range 0 to
$\EBV(\La/\Hb)$ by a $\chi^2$ fit of all the calculated line ratios and
their mean observed values (table 1 in paper I). 

\section{Theoretical and observed line profiles}
\label{s:prof}

\subsection{Theoretical line profiles}
\label{s:calc}

   The basic output of the photoionization calculations is $L_\ell(r)$, the
luminosity in a given line from a single BS at a distance $r$. The
numerical procedure for calculating $L_\ell(r)$ for a given BS wind density
structure, chemical composition and the ionizing flux spectrum is as
described in paper I.

   $L_u$, the differential line luminosity per line of sight velocity $u$, 
is given by   
\begin{eqnarray}
\label{e:Lu1} 
L_u & = & \int^{\rout}_{\rin}dr r^2 \int^{4\pi}_0 d\Omega_r 
          \int^{v_{\max}(r)}_0 dv v^2 \int^{4\pi}_0 d\Omega_v \times \nonumber \\
       &   & L_\ell(r) {\rm DF}(\bmath{r},\bmath{v})
             \delta(v\mu_r-u)\,,
\end{eqnarray}
where ${\rm DF}(\bmath{r},\bmath{v})$ is the BS distribution function and
$v_{\max}(r)$ is the maximal stellar velocity at distance r from the black
hole.  For an isotropic stellar distribution with an isotropic velocity
field, ${\rm DF}(\bmath{r},\bmath{v}) = {\rm DF}(r,v)$, $L_u$ reduces to
\begin{equation}
\label{e:Lu2} 
L_u = 8 \pi^2 \int^{r_2}_{r_1} dr r^2 L_\ell(r) 
         \int^{v_{\max}(r)}_{|u|} dv |v| {\rm DF}(r,v)\,,
\end{equation}
where the $\delta$ function is translated into the integration range
$(r_1,r_2)$, which is the sub-interval of $(\rin,\rout)$ where $|u| \le
v_{\max}(r)$. For the assumed Maxwell-Boltzmann velocity field
(equation~\ref{e:MB}) $v_{\max}(r) \rightarrow \infty$ and the line profile
is
\begin{eqnarray} 
\label{e:Lu3} 
L_u & = & \sqrt{24\pi} \int^{\rout}_{\rin} dr r^2 L_\ell(r)
      \frac{n_\star(r)}{v_0(r)}
      \exp\left[-\frac{3}{2}\left(\frac{u}{v_0(r)}\right)^2\right]\,. 
\nonumber \\
&&      
\end{eqnarray} 
Equation~\ref{e:Lu3}, together with the $L_\ell(r)$ results of the
photoionization calculation and the approximate $v_0(r)$
(equation~\ref{e:v0}), are used below to predict the line profiles of
several different BS models.


\subsection{Observed line profiles}
\label{s:data}

   The calculated line profiles are compared to a sample of 87 $\CIIIb$
profiles, 140 $\CIV$ profiles and 72 $\MgII$ profiles, for which we
measured the full line widths (FW) at 1/4, 1/2 and 3/4 of the maximum
height. This sample was compiled from several sources. 111 of the $\CIV$
lines are from an absorption line survey of 123 high luminosity AGN
\cite{SBS,SSB,SS95}, which were selected by apparent magnitude and
redshift. The QSO lie primarily in the redshift range $z \sim 1.9$ to $3.5$
and the in the absolute magnitude range $M_V \sim -32$ to $-27$.  The
continuum subtracted spectra of this sample, which were measured here, are
from Wills et al \shortcite{WBFSS}. 
The $\CIIIb$ and $\MgII$ and 29 of the $\CIV$ profiles are from a sample of
42 radio-selected QSO and 50 radio-quiet QSO, which were selected by UV
excess or slitless spectroscopy \cite{SS91}. The QSO lie primarily in the
redshift range $z \sim 0.9$ to $2.2$ and in the rest-frame monochromatic
absolute magnitude range $M(\lambda2200{\rm\AA}) \sim -30$ to $-26$. The
continuum subtracted spectra of this sample, which were measured here, are
from Brotherton et al \shortcite{BWSS}. All the continuum subtracted spectra
were kindly supplied by M. Brotherton.

\begin{figure} 
   \centering
   \epsfxsize=240pt
   \epsfbox{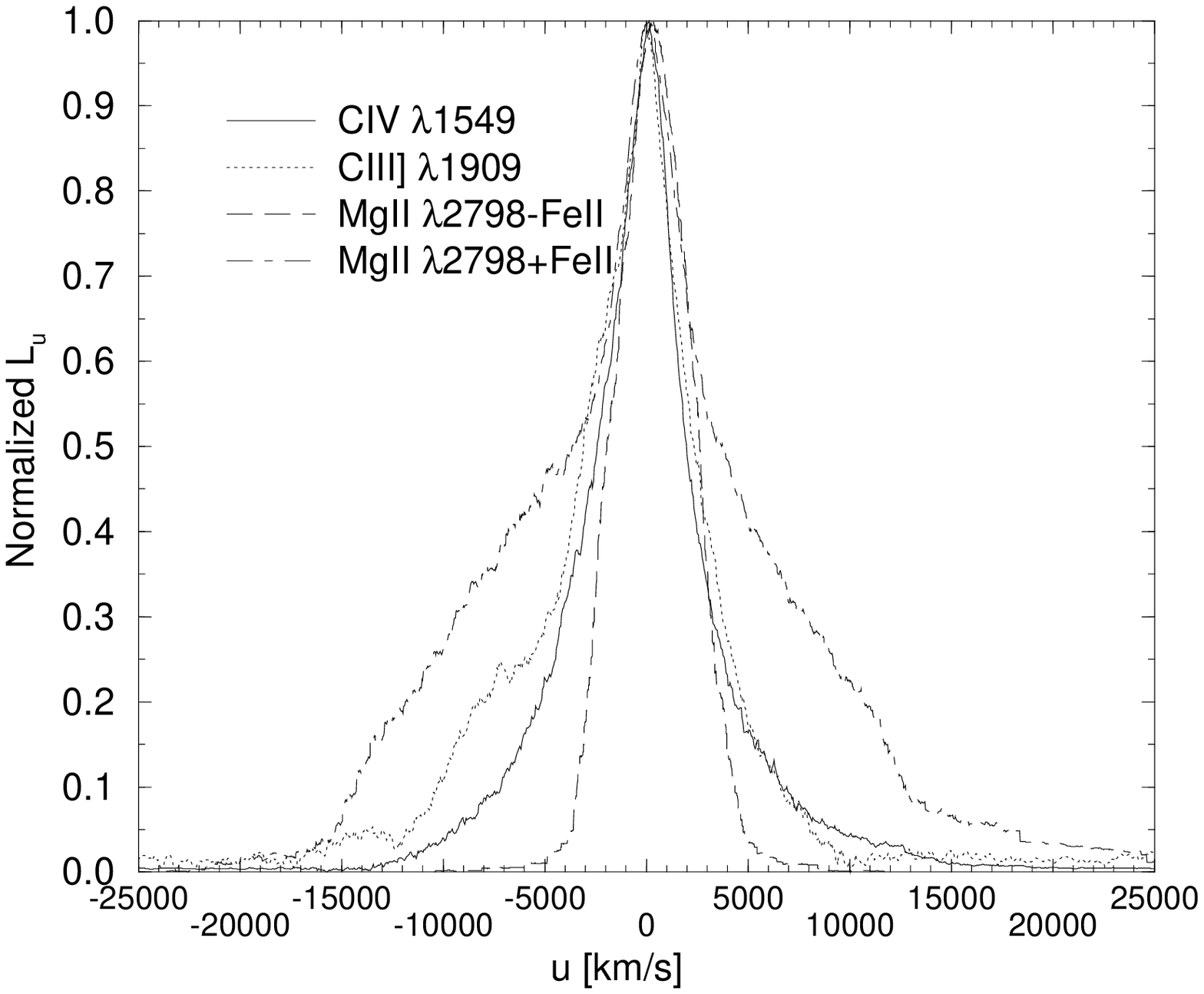}
   \caption{The average observed $\CIV$, $\CIIIb$ and $\MgII$ profiles, 
            weighted by the integrated flux. The $\MgII$ profile is shown 
            both with and without the assumed $\FeII$ contribution.}
   \label{f:avprof}
\end{figure}

   We are interested in studying the shape of the line profiles, which
requires information on the line width at several heights. A common
procedure for measuring the FW of a spectral line is to fit the profile to
a superposition of several Gaussians and measure the width of the smooth
fitted function. This procedure is also used to deblend adjacent spectral
lines from the profile. Since we are interested in the large sample
statistics rather than the properties of a specific AGN, and since the art
of multi-Gaussian fitting has a strong subjective element in it even under
optimal conditions, we decided to adopt a simpler procedure. We rely on the
fact that all lines of interest are much stronger than neighbouring lines
and do not attempt any deblending or fitting, but simply located ``by eye''
the profile peak and underlying continuum level. In cases where the blended
line was clearly visible, we adjusted the FW ``by eye'' accordingly. In
cases where the spectrum was very noisy, we smoothed it by averaging the
flux in overlapping windows of $n$ consecutive pixels (usually $n=5$).

   The measurement of the $\MgII$ FW is complicated by the fact that this
line is blended with a number of $\FeII$ lines, which add to the $\MgII$
profile wide and flat broad wings \cite{WNW}. Since it is not our intention
to do here a careful deblending of the $\MgII$ profile, we measured the
$\MgII$ twice, once with these wings and once without them. In the latter
case, only 68 profiles were measured since the others lacked distinct
wings. These two types of measurements, which we designate by $\MgIIw$ and
$\MgIIc$, respectively, represent the two extreme assumptions that either
the wings are all $\MgII$ emission or that only the line core is $\MgII$
emission. Thus, a BLR model that predicts FW(\MgII) $\ll$ FW(\MgIIc) or
FW(\MgII) $\gg$ FW(\MgIIw) is inconsistent with the observations.

   We compared our FWHM results with the Gaussian fit results of Wills et
al \shortcite{WBFSS} and Brotherton et al \shortcite{BWSS} by calculating
the mean fractional difference, $(\mbox{\rm FWHM}_{\mbox{\rm by
eye}}-\mbox{FWHM}_{\mbox{\rm gauss}}) / \mbox{FWHM}_{\mbox{\rm gauss}}$,
and its rms scatter. For the $\CIV$, $\CIIIb$, $\MgIIw$ and $\MgIIc$
profiles the mean fractional differences are $-0.009\pm0.06$,
$0.05\pm0.15$, $-0.01\pm0.1$ and $-0.26\pm0.13$, respectively. The
correspondence of the two methods for the $\CIV$ and $\MgIIw$ lines is very
good. The relatively larger over-estimation of the $\CIIIb$ FWHM is due to
the blended ${\rm Al}\,{\sc iii}\,\lambda1859$ and the big difference in
$\MgIIc$ is due to the fact that Wills et al \shortcite{WBFSS} and
Brotherton et al \shortcite{BWSS} did not deblend the $\FeII$ contribution
to the $\MgII$ profile.

\begin{figure*}
   \centering \epsfxsize=500pt
   \epsfbox{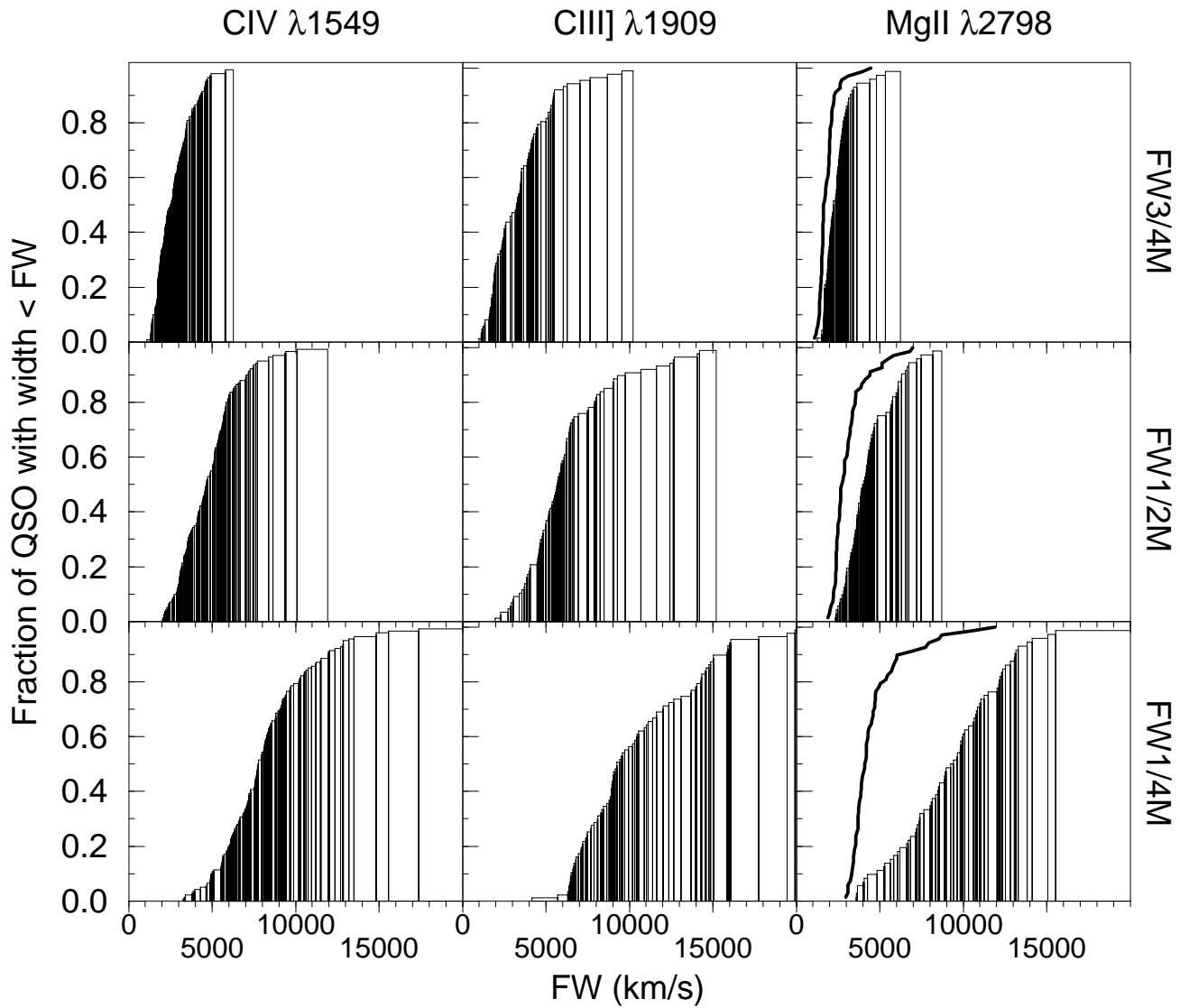} 

   \caption{The cumulative distribution functions of the observed FW at
            1/4, 1/2 and 3/4 maximum for $\CIV$, $\CIIIb$ and $\MgII$.  The
            QSO number density per FW interval is proportional to the
            density of vertical lines. The $\MgII$ cumulative FW
            distribution is based on the $\MgIIw$ measurements.  The
            $\MgIIc$ distribution is outlined by the bold line.}
   \label{f:cum}
\end{figure*}

   Figure~\ref{f:avprof} shows the sample averaged profiles, where the
profiles are weighted by their inverse integrated flux. It should be
noted that other weighing methods (e.g. by the peak flux), or other
averaging procedures (e.g. averaging the blue and red half widths at
$x$\% of the maximum, for $x = 0$ to 100\%) yield very different
results. This ambiguity indicates that the average profile is not a very
useful quantity and that the calculated profiles should be evaluated by
the cumulative distribution function or on an object to object basis.
Figure~\ref{f:cum} shows the FW cumulative distribution functions of the
three lines at 1/4, 1/2 and 3/4 maximum. 

\section{Results}
\label{s:result}

   We investigate the line profiles of the BS model by studying a
succession of three BS wind / stellar DF models, whose properties are
summarised in table~\ref{t:modelABC}. Model A is similar to
that studied in paper I. We show below that in order to reproduce the
observed line profiles and ratios, some changes in the model assumptions
are required. These changes are introduced in two steps, in models B and C.

\begin{table*}
\centering
\begin{tabular}{lccc}
\hline
& Model A & Model B & Model C \\
\hline
$\vb$                    & $5\e{3}$ cm s$^{-1}$ &$8\e{4}$ cm s$^{-1}$ &
                           $8\e{4}$ cm s$^{-1}$ \\
$\Mbh$                   & $8.0\e{7}\Mo$ & $1.9\e{8}\Mo$ & $1.9\e{8}\Mo$ \\
$t_0$                    & $3\e{8}$ yr & $10^9$ yr & $10^9$ yr \\
$\Lion$                  & $7\e{45}$ erg s$^{-1}$ & $3.6\e{44}$ erg s$^{-1}$ & 
                           $3.6\e{44}$ erg s$^{-1}$ \\
$\rin$                   & 0.001 pc & 0.001 pc & 0.001 pc \\
$\rout$                  & 1 pc & 0.25 pc & 0.25 pc \\
No. of BSs               & $1.5\e{5}$ & $4.3\e{4}$ & $4.7\e{4}$ \\
Fraction of BSs          & $4.7\e{-4}$ & $2.4\e{-3}$ & $2.6\e{-3}$ \\
$\fBS$                   & $\propto r^0$ &  $\propto r^0$ & $\propto r^{-2}$ \\
$\dot{M}_{\rm evol}$     & $0.15\Mo$ yr$^{-1}$ & $4.3\e{-2}\Mo$ yr$^{-1}$ & 
                           $4.8\e{-2}\Mo$ yr$^{-1}$ \\
$\dot{M}_{\rm BS\,coll}$ & $3.2\Mo$ yr$^{-1}$ & $0.37\Mo$ yr$^{-1}$ & 
                           $0.59\Mo$ yr$^{-1}$ \\
$\dot{M}_{\rm MS\,coll}$ & $2.8\e{-2}\Mo$ yr$^{-1}$ &$1.1\e{-2}\Mo$ yr$^{-1}$& 
                           $1.1\e{-2}\Mo$ yr$^{-1}$ \\
Electron scattering $\dot{\tau}$   
                         & $<3.8\e{-4}$  yr$^{-1}$& $<1.8\e{-4}$ yr$^{-1}$ &
                           
$<0.08$ yr$^{-1}$ \\
$\EBV$                   & 0.066 & 0.000 & 0.096 \\
$\La$ flux weighted radius   & 493 ld & 209 ld & 25 ld \\
\hline
\end{tabular}
\caption{The parameters and calculated properties of models A, B and C.
The growth rate of the electron scattering optical depth, $\dot{\tau}$, is
estimated under the assumption that all the mass accumulates in the
BLR and is neither accreted nor ejected (see paper I).}
\label{t:modelABC}
\end{table*} 

\subsection{Model A}

   We begin by considering the successful BS model of paper I (hereafter
model A, see table~\ref{t:modelABC}), which has the BS wind parameters
$\alpha = 1/2$ and $\vb = 5\e{3}$ cm/s, an AGN model of $\Mbh = 8\e{7}\Mo$
and $\Lion = 7\e{45}$ erg/s at $t_0 = 3\e{8}$ yr and $\fBS = {\rm
const}$. A free parameter of the model is the BLR size, $\rout$. As was
shown in paper I, the line spectrum depends only weakly on $\rout$ in the
range 1/3 pc to $\sim 50$ pc. This is not the case for the line width,
which decreases significantly with $\rout$. We will assume that the BLR
outer limit scales as $\rout \sim(\Lion/10^{46}{\rm erg/s})^{1/2}$ pc. This
is motivated both by the line reverberation results that suggest that the
average BLR radius is $\sim 0.1(\Lion/10^{46}{\rm erg/s})^{1/2}$pc
\cite{BNW} and by the possibility that the line emission is suppressed by
dust at the dust sublimation radius $r_{\rm sub} \simeq
0.3(\Lion/10^{46}{\rm erg/s})^{1/2}$pc \cite{NL}. We therefore set $\rout =
1$ pc for model A. 

   Figure~\ref{f:profABC} shows some profiles calculated for a BLR size
$\rout = 1$ pc.  Three properties are immediately apparent: The profiles
are very narrow (FWHM $\sim 2200$ km/s), the wings are very weak and the
different lines have almost identical profiles, unlike the observed ones
(Figure~\ref{f:avprof}). Table~\ref{t:FW} shows that only a few percent of
the objects in the sample have FWHM as narrow as those calculated, and none
have such narrow FW1/4M. We explored also the possibility of reducing
$\rout$ to 1/3 pc. This increased the $\CIV$ FW to 1877 ($\ge 29\%$ of the
sample), 2982 ($\ge14\%$) and 4422 ($\ge5\%$) km/s for the FW3/4M, FW1/2M
and FW1/4M, respectively. However, this modest increase in FW1/4M comes at
the price of an excessive BS--BS collisional mass loss rate of
$4.5\Mo$/yr. As will be discussed below, the emissivity of tidally-limited
BSs is approximately independent of $r$. Decreasing the BLR size therefore
means that the same number of BS must share a smaller volume and this leads
to an increased collision rate.

\begin{table*}
\centering
\begin{tabular}{lccccccccc}
\hline
 & \multicolumn{3}{c}{Model A} &
   \multicolumn{3}{c}{Model B} &
   \multicolumn{3}{c}{Model C} \\
FW & 
   $\CIV$ & $\CIIIb$ & $\MgII$ &
   $\CIV$ & $\CIIIb$ & $\MgII$ &
   $\CIV$ & $\CIIIb$ & $\MgII$ \\
\hline
3/4M &  
    1337  &  1407    &  1252   & 
    2398  &  2426    &  2366   &
    2854  &  2958    &  2729   \\
\%     &
     (4)  &   (8)    &   (1/5)   &       
    (49)  &  (38)    &  (54/91)  &
    (61)  &  (47)    &  (76/96)  \\
1/2M &
    2090  &  2202    &  1952   & 
    3734  &  3789    &  3709   &
    4561  &  4782    &  4319   \\
\%    &
     (2)  &   (2)    &   (0/3)   &       
    (33)  &  (15)    &  (41/84)  &
    (48)  &  (31)    &  (62/89)  \\
1/4M &
    2988  &   3164   &  2783   & 
    5372  &   5464   &  5813   &
    6928  &   7344   &  6424   \\
\%    &
     (0)  &    (0)   &   (0/0)   &       
    (12)  &    (2)   &  (16/86)  &
    (32)  &   (23)   &  (20/90)  \\
\hline
\end{tabular}
\caption{Comparison of the calculated FW to the observed cumulative FW
         distribution. The FW are in km/s. The numbers in brackets below
         each FW entry are the percentage of objects in the sample with FW
         less or equal to the calculated one. For the $\MgII$, the
         percentage on the left is based on the $\MgIIw$ FW measurements,
         and that on the right on the $\MgIIc$ FW measurements.}
\label{t:FW}
\end{table*}

\subsection{Model B}

   A possible remedy for the width problem is to model a more evolved AGN
with a more massive black hole. The MCD models indicate, that apart for a
very short initial epoch of Eddington limited accretion, the luminosity
decreases with time as the available mass is exhausted (MCD figure 10). The
resulting decrease in $\rout$ also helps to broaden the profiles. We
therefore consider an AGN model at $t_0 = 1$ Gyr, when $\Mbh =
1.9\e{8}\Mo$, $\Lion = 3.6\e{44}$ erg/s and $\rout = 1/4$ pc (hereafter
model B, see table~\ref{t:modelABC}). As is explained in the appendix,
changing $\Mbh$ and $\Lion$ from the values of model A to those of model B
worsens the fit of the calculated line ratios to the observed ones, unless
accompanied by a suitable modification of the wind parameters. In paper I,
we systematically searched the BS wind parameter space for the values that
best reproduce the observed line ratios. Here, we use instead the scaling
prescription given in the appendix, which shows that the line spectrum of
model A can be approximately reproduced by modifying the wind velocity to
$\vb = 8\e{4}$ cm/s. For the continuum spectrum we use the same spectral
distribution as used for model A, scaled down to the lower $\Lion$ of model
B.

\begin{figure}
   \epsfxsize=240pt
   \epsfbox{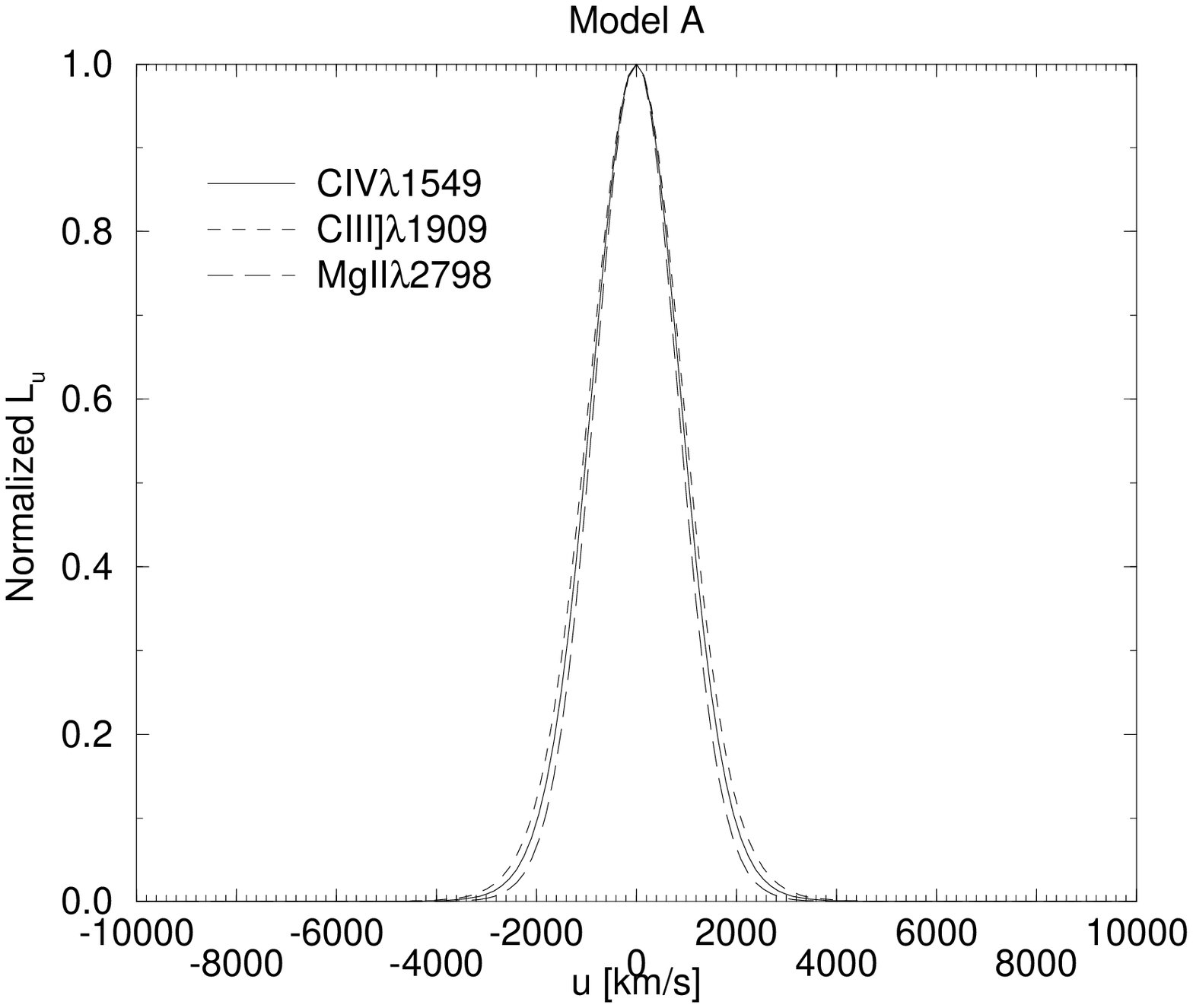}
   \epsfxsize=240pt
   \epsfbox{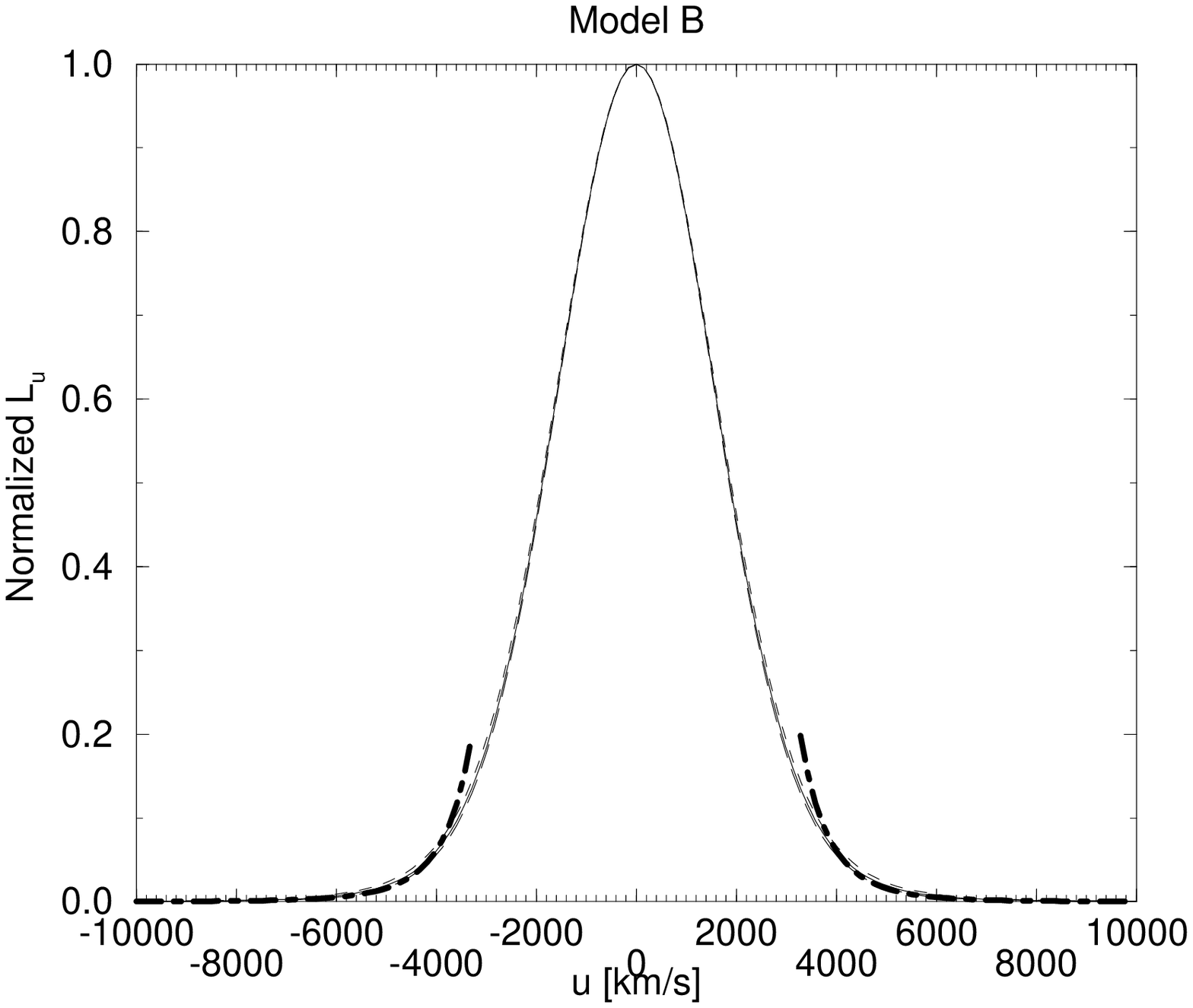}
   \epsfxsize=240pt
   \epsfbox{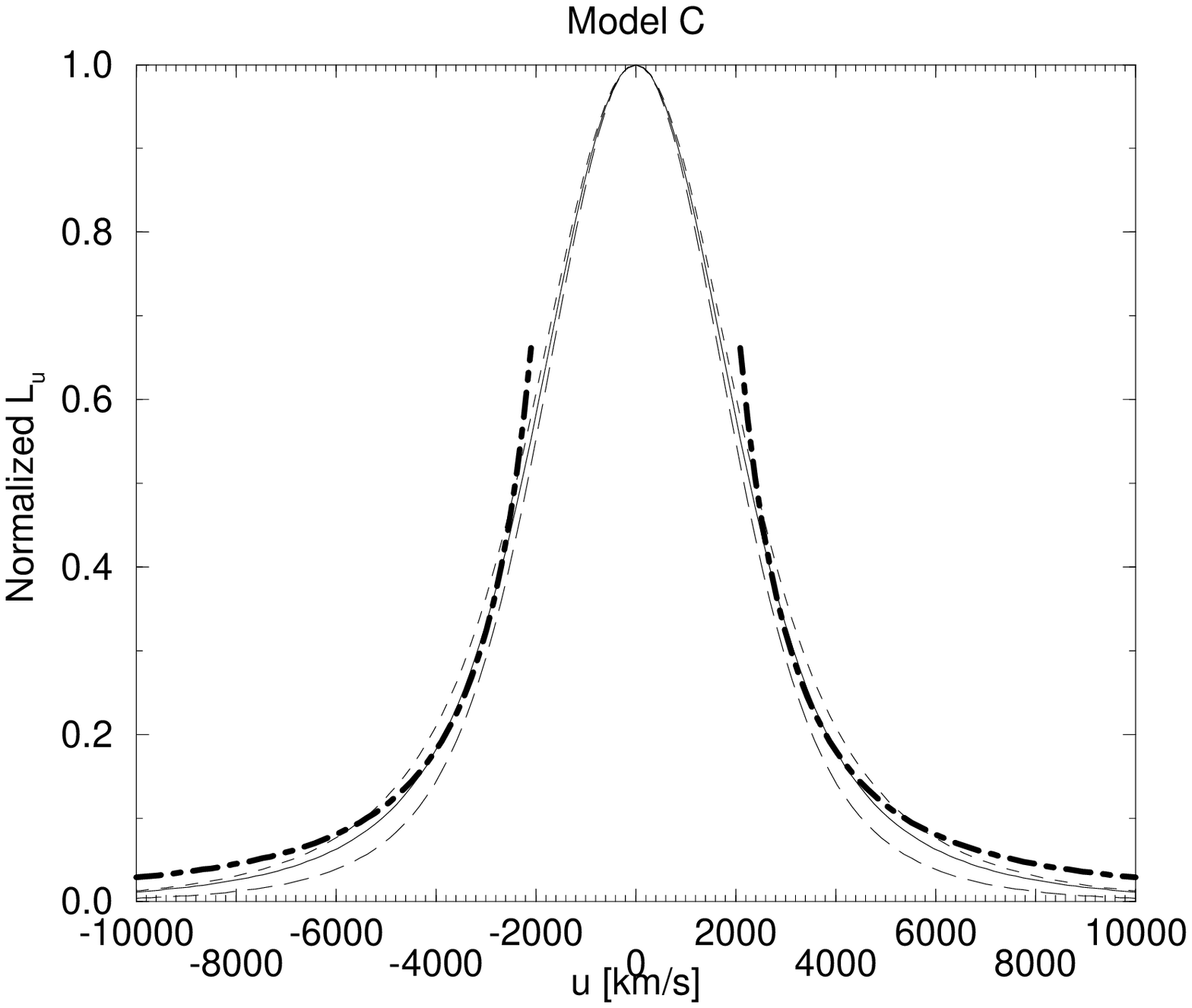}
   \caption{The theoretical $\CIV$, $\CIIIb$ and 
            $\MgII$ profiles for models A, B and C.
            The model B numerical profiles are fitted to the 
            analytical approximation $L_u \propto u^{-6}$
            (bold dash-dotted line).
	    The model C numerical profiles are fitted to the
            analytical approximation $L_u \propto u^{-2}$
            (bold dash-dotted line).
           } 
    \label{f:profABC}   
\end{figure}

  The calculated BS--BS collisional mass loss rate of model B is only
$0.4\Mo$/yr, in spite of the small BLR size. This reflects the fact that
most of the mass loss is from collisions of the winds, and that the BS wind
density is proportional to $\dMw/\vb$ (equation~\ref{e:rmass}). Since $\Rt$
is a weak function of $\Mbh$, the BS wind mass in model B is $\sim 16$
times smaller than that of model A, and so are the respective mass loss
rates.  Figure~\ref{f:profABC} and table~\ref{t:FW} show that the profiles
of model B are indeed wider than those of model A, but the profiles of the
three lines are almost identical and the wings are too weak.

   So far, we assumed that the fraction of BSs in the central cluster,
$\fBS$, is independent of $r$. There are, however, good reasons to modify
this approach, both from theoretical considerations and from careful
examination of the line wings. As noted by Penston, Croft, Basi \& Fuller
\shortcite{PCBF}, the far wings of many observed profiles are well fitted
by an inverse quadratic form of $F_\lambda \propto 1/(\lambda-\lambda_0)^2$
and show that this is consistent with line emission from an ensemble of
clouds of constant cross section, moving on parabolic orbits. Here, we
extend their analysis to the BSs and show that the discrepancy between the
observed wing shapes and those of model B can be explained in terms of the
line emissivity from marginally bound BSs.

   Let us assume for simplicity that the Maxwellian DF
(equation~\ref{e:MB}) can be approximated by
\begin{equation}
DF(r,v) = \fBS(r)\ns(r)\frac{\delta[v-v_0(r)]}{4\pi v^2}\,,
\end{equation}
where $v_0^2 = 2G\Mbh/r$. It then follows from equation~\ref{e:Lu2}
that
\begin{equation}
L_u = 2\pi \int_{r_1}^{r_2} dr r^2 L_\ell(r)\fBS(r)\ns(r)v_0^{-1}(r)\,,
\end{equation}
where the upper integration limit is defined by $r_2 = \min(
2G\Mbh/u^2,\rout)$ and $r_2 \gg r_1$. As explained earlier, marginally bound
stars have $v_0(r) \propto r^{-1/2}$. Assume that $L_\ell(r) \propto
r^{-a}$, $\fBS(r) \propto r^{-b}$ and $\ns(r) \propto r^{-c}$. It then
follows that
\begin{equation}
L_u \propto r_2^{7/2-(a+b+c)}\,.
\end{equation}
At the profile wings $u$ is large enough so that $r_2 < \rout$, $r
\propto u^{-2}$ and the profile shape is  
\begin{equation}
\label{e:Luexp}
L_u \propto u^{-7+2(a+b+c)}\,.
\end{equation}

\begin{figure}
   \centering \epsfxsize=240pt
   \epsfbox{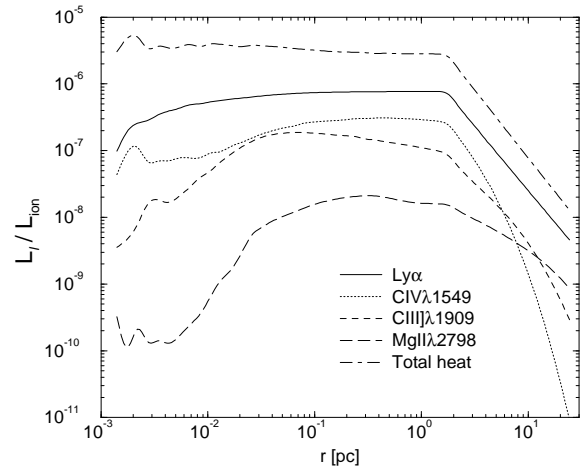} 

   \caption{The line emissivity of a single BS, $L_\ell$, for models B and
   C in various lines, as function of the distance from the black hole. The
   total heat includes all the reprocessed radiation emitted in lines and
   continuum radiation.}
\label{f:Lstar}
\end{figure}

   The tidally-limited BSs are dense and optically thick enough, so that
their total emissivity is roughly proportional to the product of their
geometrical cross-section, $\pi R^2_{\rm tidal} \propto r^2$
(equation~\ref{e:rt}) and the central irradiating flux, $F \propto r^{-2}$,
and is therefore constant. Figure~\ref{f:Lstar} shows the BS line and total
emissivity of model B. The total emitted heat (including all lines and
diffuse continua) is indeed very nearly constant .This, however is not true
for the individual lines (Fig.~\ref{f:Lstar}), whose emissivity depends on
the gas density, ionization level and optical depth. Nevertheless, let us
make the approximation that $L_\ell \sim {\rm const.}$ ($a = 0$) and assume
also that $\fBS = {\rm const.}$ ($b = 0$). Marginally bound stars have
$\ns(r) \propto r^{-1/2}$ $(c = -1/2)$ (e.g. Penston et al 1990). The
profile wing is therefore expected to fall off sharply as $L_u \propto
u^{-6}$. Figure~\ref{f:profABC} shows that the calculated profiles of model
B are well fitted by this expression at high $u$, thereby also justifying
the approximation of the DF by the $\delta$ function.

\subsection{Model C}

\subsubsection{Line profiles}

   Equation~\ref{e:Luexp} and the observed profiles suggest that in
order to reproduce the observed inverse quadratic wings, it is
necessary to modify some of the model assumptions, so that $a+b+c =
5/2$. Changing the BS emissivity by modifying the wind structure will
change the emission line spectrum, as shown in paper I. The $r$
dependence of the stellar DF is based on a well established
result. The only remaining free parameter is $\fBS$. If we assume
$\fBS \propto r^{-2}$ then the profile wings will have the required
shape\footnote{$\fBS \propto r^{-2}$ corresponds to a differential
geometric covering factor that falls off as $\propto r^{-1/2}$, since
$\Rt \propto r$ and $\ns \propto r^{-1/2}$.}.

   Model C (table~\ref{t:modelABC}) has the same parameters as model B but
with $\fBS \propto r^{-2}$ (equation~\ref{e:fBSr2}). Figure~\ref{f:profABC}
and table~\ref{t:FW} show that with this choice, the calculated FWHM are
representative of the observed sample. However, the percentiles decrease
from the FW3/4M to FW1/2M to FW1/4M, indicating that the calculated wings
are still somewhat weaker than the observed ones. The calculated $\MgII$
FW3/4M is now in fact wider than more than 76\% percent of the sample
profiles.  The fit to $L_u \propto u^{-2}$, shown in figure~\ref{f:profABC}
is not perfect. This is due to the fact that $L_\ell$ of those lines is not
constant. We have examined the fit to the calculated $\La$ profile and
found that for this line, whose emissivity is closer to being constant with
$r$, The fit of the far wings to $L_u \propto u^{-2}$ is very good.  

\begin{figure*}
   \centering
   \epsfxsize=450pt
   \epsfbox{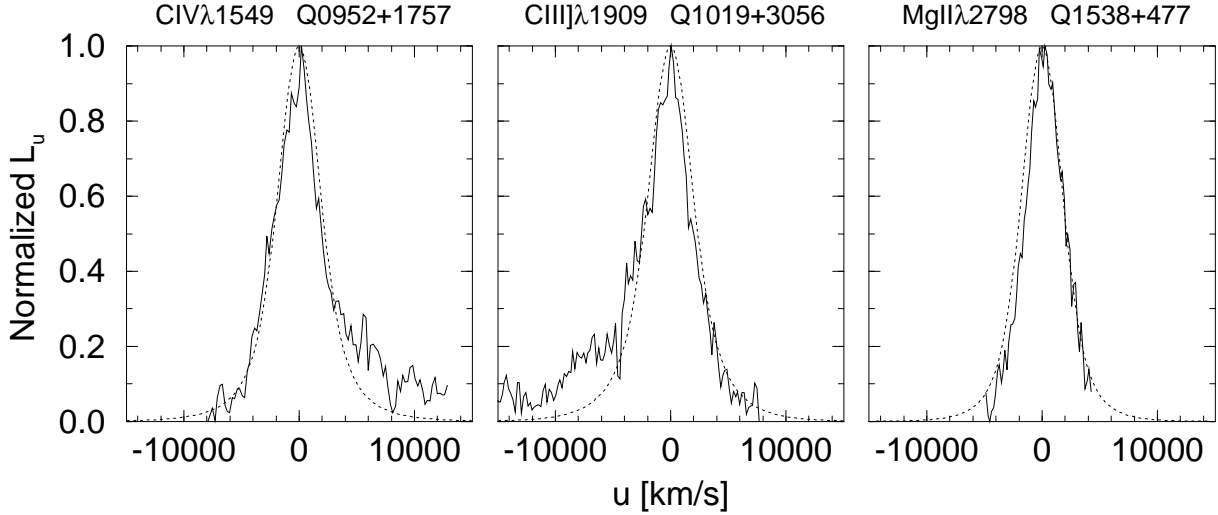}
   \caption{Examples of observed profiles (solid line) which are well 
	    fitted by model C (dotted line).
            } 
\label{f:fit3}   
\end{figure*}

   The FWHM ratios of the calculated lines of model C are $\CIIIb/\CIV =
1.05$, $\MgII/\CIV = 0.95$ and $\MgII/\CIIIb = 0.90$. These values can be
compared to the observed values quoted in Brotherton et al (1994),
$1.21\pm0.04$, $0.97\pm0.06$ and $0.85\pm0.05$, respectively. It is
apparent that the model $\CIIIb$ tends to be narrower than the observed
one, as is also indicated by the percentages in
table~\ref{t:FW}. Figure~\ref{f:fit3} demonstrates that there are AGN which
have profiles that are very similar to those of model C.


\subsubsection{Line ratios}

\begin{figure}
   \centering
   \epsfxsize=240pt
   \epsfbox{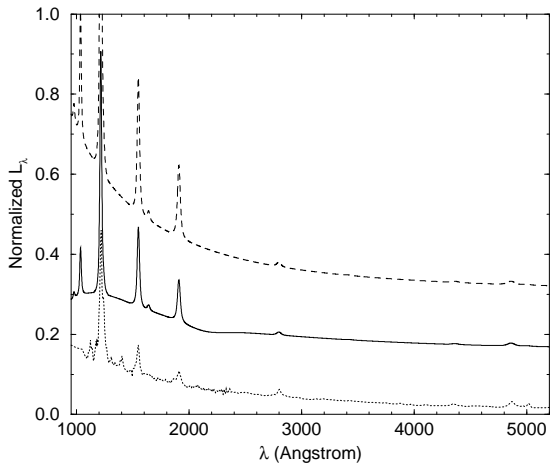}
   \caption{The unreddened (dashed line) and reddened, $\EBV = 0.096$, (solid
            line) spectrum of model C compared with an average observed
            QSO spectrum (dotted line).}

   \label{f:compare}   
\end{figure}

\begin{figure} 
   \epsfxsize=240pt
   \epsfbox{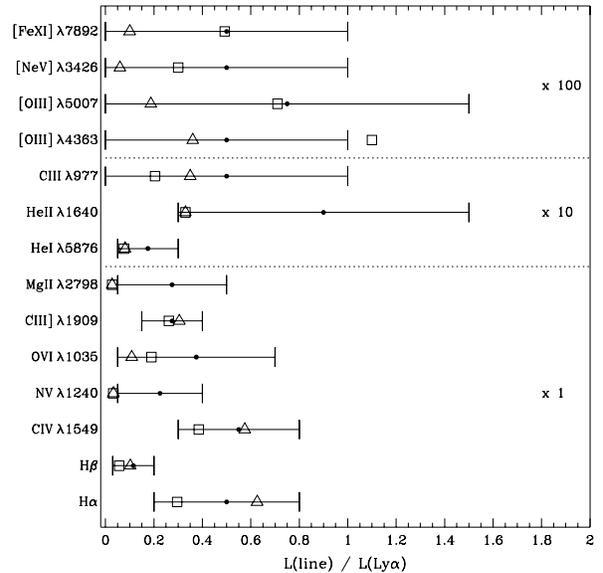}
   \caption{The integrated line ratios of models A~($\triangle $) and
            C~($\Box$), expressed relative to the $\La$ luminosity and
            compared to the observed range (paper I, table 1). The mean
            observed ratios are indicated by the black dot and the ranges
            by the error bars. $\EBV({\rm A}) = 0.066$ and $\rout({\rm A})
            = 1$ pc.  $\EBV({\rm C}) = 0.096$ and $\rout({\rm C}) = 1/4$
            pc.}  
\label{f:ratios}
\end{figure}

   Having established a good agreement between the observed and
calculated profiles, we now further test model C by returning to the
issues studied in paper I, namely the line ratios and the required
number of BSs. The integrated line spectra of models A and C are
compared to the mean observed one (paper I, table 1) in
figure~\ref{f:ratios}. (Note the similarity between the line ratios of
models A and C, which demonstrates that the scaling prescription of
the appendix works). The $\MgII$ and $\NV$ deficiencies, which are
discussed in paper I, remain a problem. Model C has a slightly high
$\bOIIIbA$ emission, which is due to the fact that the scaling only
deals with the tidally-limited BSs. At large $r$ the BS are
mass-limited. The lower wind gas density of model C implies a larger
$\Rm$ and consequently a lower density, a larger $U(\Rm)$ and a
stronger emission of forbidden lines. The unreddened and reddened
spectra of model C are compared to a mean QSO spectrum in
figure~\ref{f:compare}.

   Model C is quite successful in reproducing the BLR emission
properties. The line ratios, line profiles and BLR size, as estimated by the
flux weighted radius of the $\La$. The total mass loss rate is consistent
with the theoretical estimates of the accretion rate and the resulting
electron scattering optical depth growth is consistent with the
observations of rapid X-ray variability.


\section{Discussion and conclusions}
\label{s:discuss}

   The results of paper I indicate that in order to reproduce the
integrated emission line spectrum, the BSs should have dense, decelerating
winds, whose boundary is not fixed by comptonization. The BLR line profiles
are another observed property that the BS model has to reproduce. We find
that this imposes additional constraints on the model.

   The FW statistics of the observed profiles indicate that the high
$\LtoM$ models that were successful in reproducing the emission line
spectrum have too narrow FWs. Observed AGN with such narrow broad lines
constitute only a marginal percentage of the sample. Another potential
problem with these models is the rather high collisional mass loss rate.
The narrow calculated FWs indicate that either $\Mbh$ should be larger
and/or the BLR size smaller. The MCD models of the AGN evolution link the
increase in the black hole mass with a decrease in the luminosity (in the
post-Eddington limited accretion epoch). The change in the luminosity
affects the model properties in two different ways. First, it may decrease
the BLR size, since it is natural, although not necessary, to assume that
the luminosity determines the BLR size by creating a dust-free inner region
which can emit lines. Second, the reduced BLR size and increased black hole
mass make the tidally-limited BSs the main contributors to the line
emission. We find that, in order to obtain the observed line spectrum from
tidally-limited BSs, it is necessary to scale $\dMw/\vb$ in proportion to
$\Lion/\sqrt{\Mbh}$ (see the appendix). This in turn reduces the wind mass
and allows the BSs to survive the increased collisional rate in the
smaller, faster moving BLR. The lower $\LtoM$ models have much wider
profiles, which are consistent with the observations. However the far wings
are very weak and the profiles of the different species are very similar to
each other, contrary to the observed situation.

   Although the profile widths of model A are much smaller than those of
the sample presented here, we note that there exists a sub-population of
narrow line Seyfert 1 galaxies, which are characterized by soft X-ray
spectra and comprise about 10\% of optically selected Seyfert 1s and
15--50\% of soft X-rays selected Seyfert 1s \cite{BBF,Laor}. Such AGN have
FWHM($\Hb$) in the range $\sim$ 500--1500 km/s, which is similar to, or
even narrower than the FWHM of model A.

   The idea that the inverse quadratic far wings, which are observed in
many profiles, are linked to parabolic orbits and to $r^{-2}$ emissivity
dependence, was discussed by several authors. The dense and deep potential
well in the AGN offers a `natural selection' process which favors such
orbits for the BSs, since all the tightly bound BSs are destroyed by
collisions or tidal disruptions. The analytic work of Penston et al
\shortcite{PCBF} assumed strictly parabolic orbits and a cloud line
emissivity that falls off as $r^{-2}$. Our numeric results confirm that
their analysis also holds for a distribution of binding energies around
$E=0$. However, unlike their assumed cloud emissivity, the total line
emissivity from a single BS is nearly constant in $r$. If we assume that
the BSs are a fixed proportion of the normal stellar population, the
resulting line profiles are core dominated with very weak wings. This
situation can be corrected by relaxing this assumption and setting $\fBS
\propto r^{-2}$. The resulting profiles are only approximately inverse
quadratic, because the emissivity of the individual lines is not
constant in $r$. The BS population of this model is more concentrated
towards the central high velocity region. As a consequence, the
differences in the emissivity of the various lines begin to manifest
itself in profile differences. The idea that the BS fraction decreases
with distance from the black hole is a natural one since such stars
are not seen in normal stellar environments and their existence must
therefore be due to the special conditions in the AGN. 
 
   It is tempting to interpret the $r^{-2}$ dependence as indicating
that the local photon or particle flux is connected to the creation of
BSs.  However, taken at face value, $\fBS\propto r^{-2}$ means only
that the BS {\em number} is a function of distance from the black
hole. It implies nothing about the distance dependence of the BS {\em
properties}. In the simplified picture presented in this paper, normal
stars approach their perigalacton on a near-parabolic orbit, and due
to some yet unspecified physical mechanism, a small fraction of them
expands to the BS state at various distances from the black hole. If
the bloating process occurs in a short time relative to the BLR
crossing time (e.g. a collisional merger between two stars), $\fBS(r)$
can be interpreted as the probability for this to happen to a given
star at a given radius. Alternatively, the bloating may be a gradual
process. In this case, the simplified two-component stellar population
that was used here may be thought of as an approximation, where all
stars bloated beyond a certain threshold are represented by a single
type of ``effective'' BS, and the rest by a solar type star. $\fBS(r)$
is then the fraction of stars that are bloated beyond this threshold
at a given radius.

   Although the modified BS model succeeds in reproducing the FWs and wide
wings to a fair degree, there remain several problems. The calculated
profile wings are still weaker than the observed ones. This may be due to
the fact that the cores of the observed profiles have a blended
contribution from the narrow line region. In addition, the differences
between the lines are not as distinct as in the mean observed profiles. A
reliable estimate of this discrepancy is possible here only for the
$\CIIIb$/$\CIV$ FWHM ratio. The $\MgII$ line is problematic, both in the
model, which badly under-estimates its strength, and in the observed
spectra, where it is very difficult to deblend. These discrepancies may
indicate that $\fBS$ falls off even faster than $r^{-2}$.  

   The question whether BSs can produce the extremely broad profiles
that are seen in some objects was not investigated here. This is part
of the larger question of whether the BS model has the flexibility to
explain the entire range of observed AGN spectra, which will be
addressed in a future paper. Here, we limit ourselves to the problem
of reproducing the profiles of a typical AGN, where we define typical
profiles as those that have the median FW and inverse quadratic
wings. We also did not address the issues of profile asymmetry and
profile shifts relative to the systemic one. In this context, it is
interesting to note that there have been phenomenological attempts to
decompose the BLR emission spectrum into several components, and in
particular to assign the line shifts and profile asymmetry to a very
broad component
\cite{FHFC,WBFSS,BWSS,BWFS,Betal}. In such a scenario, the BSs can be
identified with the symmetric, unshifted, intermediate width
component.

   The conclusion that the BSs are likely to be tidally-limited and
optically thick has an interesting consequence. For a given BLR covering
factor, $C_F$, (usually $\sim 0.1$) the total number of BSs, $N_{\rm BS}$,
is
\begin{equation} 
N_{\rm BS} = 
\frac{4C_F}{X^2_{\rm tidal}}\left(\frac{\Mbh}{\MBS}\right)^{2/3}\,.
\end{equation}
This explains why in all our models, $N_{\rm BS} \sim 5\e{4}$. It also implies
that if $X_{\rm tidal}$ and $\MBS$ are similar in different AGN, $N_{\rm
BS}/C_F$ is a measure of the black hole mass.

  To summarize, our attempts to model the BLR line profiles with BSs lead
us to the following conclusions.
\begin{enumerate}
\item
   The BSs can reproduce the typical full widths of observed profiles if
   $\LtoM$ is low enough (e.g. $\Mbh = 2\e{9}\Mo$ and $\Lion = 3.6\e{44}$
   erg/s) and an external cutoff mechanism limits the size of the BLR.
\item
   Due to the small size of the BLR, the most important effect limiting the
   size of the BSs is tidal disruption by the black hole.
\item
   In order to reconcile the observed similarity of AGN line spectra over a
   large range of continuum luminosities with the idea of tidally limited
   BSs, it is necessary to assume that the wind properties (e.g. mass loss
   rate or wind velocity) are correlated with $\Mbh$ and $\Lion$.
\item 
   The BSs can produce the observed wide, inverse quadratic far wings if
   their fraction in the stellar population falls off roughly as
   $r^{-2}$. Although this is a highly concentrated distribution, the
   collisional mass loss is small enough to be consistent with the
   observations and BS lifetimes. However, the calculated profiles tend to
   have weaker wings than the observed ones and the differences between the
   profiles of the different species are somewhat smaller than those actually
   observed.
\end{enumerate}

{\bf Acknowledgments}\hfill\\

   We are very grateful to Mike Brotherton for the reduced spectra, to
Mingsheng Han for the spectrum reduction program {\sc specl} and to Brian
Murphy for the stellar velocity data. This research was supported by the
Israel Science Foundation administered by the Israel Academy of Sciences
and Humanities and by the Jack Adler chair of Extragalactic Astronomy at
Tel Aviv University.

\section*{Appendix: $\LtoM$ scaling of the wind model}
\setcounter{equation}{0}
\renewcommand{\theequation}{A\arabic{equation}}
\label{s:scale}

   In paper I we demonstrated that BSs with dense and slow winds reproduce
the correct line ratios for one specific AGN model. The wind boundary of
such BSs is either tidally- or mass-limited. Unlike the Comptonization
boundary, $\Rc$, which establishes itself at a constant $U$ irrespective of
$\Lion$ or $\Mbh$, $N(\Rt)$ depends on $\Mbh$, but not on $\Lion$ and
therefore $U(\Rt)$ depends on both $\Lion$ and $\Mbh$. Similarly, $N(\Rm)$
depends on the wind properties, but not on $\Lion$ and therefore $U(\Rm)$
depends on $\Lion$. As a result, the emission line spectrum of tidally- or
mass-limited BSs with given wind parameters depends on the specific choice
of $\Lion$ and $\Mbh$. In paper I we assumed for simplicity that all the
wind properties, except the wind size, are independent of the external
environment. It is obvious that this assumption is inconsistent with
tidally-limited BS, since AGN of very different luminosities are observed
to have very similar line ratios. This leads us to hypothesize the
existence of some unspecified physical mechanism that affects the
properties of the BSs in a way that makes the emission line spectrum
insensitive to changes in $\Lion$ and $\Mbh$. The results of
section~\ref{s:result} indicate that in order to reproduce the observed
line widths, the AGN $\LtoM$ ratio should be smaller than that assumed in
paper I. In this case the line emission originates closer to the black hole
and most of it is from tidally-limited BSs. We will therefore concentrate
on the scaling properties of tidally-limited wind models.

   Let $\Lion^\prime = \lambda \Lion$ and $\Mbh^\prime = \mu \Mbh$ describe
the transformation from an AGN model with $\Lion$ and $\Mbh$ to one with
$\Lion^\prime$ and $\Mbh^\prime$. Equations~\ref{e:cont}, \ref{e:plawv} and
\ref{e:rt} and the relation $U \propto \Lion/r^2N$ yield
\begin{equation}
\label{e:U}
U(r) \propto \Rb^\alpha\MBS^{(2-\alpha)/3}X_{\rm tidal}^{2-\alpha}
          r^{-\alpha}
          \frac{\vb}{\dMw}\Lion\Mbh^{(\alpha-2)/3} \,.
\end{equation} 
where we approximate the BS mass, $\MBS$, by $1\Mo$, the maximal mass in
the BS (star and wind). It then follows that in the transformed AGN model
\begin{equation}
\label{e:Uprime}
U^\prime(r) = \lambda \mu^{(\alpha-2)/3}U(r)\,.
\end{equation}
$U$ changes with $r$, and therefore the integrated line spectrum reflects a
weighted sum over a range of $U$ values, weighted by the BS DF. The simplest
option for making the integrated spectrum constant in $\lambda$ and $\mu$
is to require that $U^\prime(r) = U(r)$. A change in the black hole mass
and luminosity imply also a change in the stellar DF and the BLR size
(i.e. the integration limits). We are assuming here that these changes have
only a small effect on the integrated spectrum. This is indeed the case in
the examples which are presented in section~\ref{s:result} (cf
Fig.~\ref{f:ratios}).

   In principle, any of the free parameters in equation~\ref{e:U} can be
varied to compensate for the scaling factor in equation~\ref{e:Uprime}. We
will assume for simplicity that
\begin{equation} 
\left(\frac{\dMw}{\vb}\right)^\prime = 
\lambda\mu^{(\alpha-2)/3}\left(\frac{\dMw}{\vb}\right)\,.
\end{equation}
For the free-fall, $\alpha = 1/2$, wind models which we study here, this
reduces to
\begin{equation}
\label{e:Mwv1}
\left(\frac{\dMw}{\vb}\right)^\prime = 
\frac{\lambda}{\sqrt{\mu}}\left(\frac{\dMw}{\vb}\right)\,.
\end{equation} 
If all AGN emit at some fixed percentage of the Eddington luminosity,
the scaling law is further reduced to
\begin{equation}
\label{e:Mwv2}
\left(\frac{\dMw}{\vb}\right)^\prime 
\propto \sqrt{\lambda}\left(\frac{\dMw}{\vb}\right)
\propto \sqrt{\mu}\left(\frac{\dMw}{\vb}\right)\,.
\end{equation} 

   The physical interpretation of these scaling laws is beyond the
scope of this work. Nevertheless, we note that it so happens that the
phenomenological scaling laws (equations~\ref{e:Mwv1} and
\ref{e:Mwv2}) take on a particularly simple and suggestive form in the
relevant case of tidally-limited and free-falling wind flows. While
this may be a clue to the physics of the winds, it should be noted
that a simple interpretation of these scaling laws in terms of the
ionizing flux or tidal forces runs into an inconsistency with our
simplified wind structure, which has no $r$ dependence.

   Unlike a systematic search in parameter space, this scaling prescription
is not guaranteed to yield the best fit results.  However, the calculations
in paper I indicate that the integrated line spectrum is a strong and
smooth function of the wind parameters. Small changes in the best fit
values of the mass loss rate or the wind velocity considerably worsen the
fit to the line spectrum. This implies that the scaling solution, or a
solution very similar to it, is likely to be the best solution in the
region of parameter space that is considered in this work.


\begin{thebibliography}{99}
\bibitem[\protect\citename{Alexander \& Netzer\ }1994]{AN}
   Alexander T,. Netzer H., 1994, MNRAS, 270, 781 (Paper I)
\bibitem[\protect\citename{Bahcall \& Wolf\ }1976]{BW}
   Bahcall J. N., Wolf R. A., 1976, ApJ, 209, 214
\bibitem[\protect\citename{Baldwin et al.\ }1995]{Betal}
   Baldwin J., A., et al, 1995, ApJ, in press
\bibitem[\protect\citename{Binney \& Tremaine\ }1987]{BT} 
   Binney J., Tremaine S., 1987, Galactic Dynamics.
   Princeton University Press, Princeton, N.J.
\bibitem[\protect\citename{Boller, Brandt \& Fink\ }1995]{BBF}
   Boller Th., Brandt W. N., Fink H., 1995, A\&A, in press 
\bibitem[\protect\citename{Brotherton et al.\ }1994b]{BWFS}
   Brotherton M. S., Wills B. J., Francis P. J., Steidel C. C.,, 
   1994, ApJ, 430, 495 
\bibitem[\protect\citename{Brotherton et al.\ }1994a]{BWSS}
   Brotherton M. S., Wills B. J., Steidel C. C., Sargent W. L. W., 
   1994, ApJ, 423, 131 
\bibitem[\protect\citename{Cohn \& Kulsrud\ }1978]{CK}
   Cohn H., Kulsrud R. M., 1978, ApJ, 226, 1087
\bibitem[\protect\citename{David, Durisen \& Cohn\ }1987]{DDC} 
   David L., P.,  Durisen R. H., Cohn H. N., 1987, ApJ, 313, 556
\bibitem[\protect\citename{Edwards\ }1980]{Edwards} 
   Edwards A. C., 1980, MNRAS, 190, 757
\bibitem[\protect\citename{Francis et al.\ }1992]{FHFC} 
   Francis P., J., Hewett P., C., Foltz C., B., Chaffee F., H., 
   1992, ApJ, 398, 476
\bibitem[\protect\citename{Hartquist et al.\ }1995]{HDDRWW} 
   Hartquist T. W., Durisen R. H., Dyson J. E., Rawlings J. M. C., 
   Williams D. A., Williams R. J. R., 1995, ApJ, 453, 77
\bibitem[\protect\citename{Kazanas\ }1989]{Kazanas} 
   Kazanas D., 1989, ApJ, 347, 74
\bibitem[\protect\citename{Laor et al\ }1994]{Laor}
   Laor A., Fiore F., Elvis M., Wilkes B. J., McDowell J. C., 
   1994, ApJ, 435, 611 
\bibitem[\protect\citename{MacAlpine\ }1985]{MacAlpine} 
   MacAlpine G. M., 1985, 
   in Miller J., ed,
   Astrophysics of Active Galaxies and Quasi-Stellar Objects.
   University Science books, Mill Valley, CA, p. 259
\bibitem[\protect\citename{Maoz et al.\ }1991]{Maoz} 
   Maoz D. et al,  1991, ApJ, 367, 493
\bibitem[\protect\citename{Mathews\ }1983]{Mathews} 
   Mathews W. G., 1983, ApJ, 272, 390
\bibitem[\protect\citename{Murphy, Cohn \& Durisen\ }1991]{MCD} 
   Murphy B. W., Cohn H. N., Durisen R. H., 1991, ApJ, 370, 60 (MCD)
\bibitem[\protect\citename{Netzer\ }1990]{BNW}
   Netzer H., 1990, in  
   Blandford R. D., Netzer H., Woltjer L.,
   Active Galactic Nuclei. Springer-Verlag, Berlin, pp. 67--134
\bibitem[\protect\citename{Netzer \& Laor\ }1993]{NL} 
   Netzer H., Laor A., 1993, ApJ, 404, L51
\bibitem[\protect\citename{Penston\ }1988]{Penston} 
   Penston M. V., 1988, MNRAS, 233, 601
\bibitem[\protect\citename{Penston, Croft, Basu \& Fuller\ }1990]{PCBF} 
   Penston M. V., Croft S., Basu D., Fuller N., 1990, MNRAS, 244, 357
\bibitem[\protect\citename{Sargent, Boksenberg \& Steidel\ }1988]{SBS}
   Sargent W. L. W., Boksenberg A., Steidel C. C., 1988, ApJS, 68, 539
\bibitem[\protect\citename{Sargent, Steidel \& Boksenberg\ }1989]{SSB}
   Sargent W. L. W., Steidel C. C., Boksenberg A., 1989, ApJS, 69, 703
\bibitem[\protect\citename{Scoville \& Norman\ }1988]{SN} 
   Scoville N., Norman C., 1988, ApJ, 332, 163
\bibitem[\protect\citename{Scoville \& Norman\ }1995]{SN2} 
   Scoville N., Norman C., 1995, ApJ, 451, 510
\bibitem[\protect\citename{Steidel \& Sargent\ }1991]{SS91}
   Steidel C. C., Sargent W. L. W., 1991, ApJ, 382, 433
\bibitem[\protect\citename{Sargent \& Steidel\ }1995]{SS95}
   Steidel C. C., Sargent W. L. W., 1995, in preparation
\bibitem[\protect\citename{Wanders et al.\ }1995]{Wanders}
   Wanders I. et al,, 1995, ApJ, 453, L87
\bibitem[\protect\citename{Wills et al.\ }1993]{WBFSS}
   Wills B. J., Brotherton M. S., Fang D., Steidel C. C., Sargent W. L. W.,
   1993, ApJ, 415, 563
\bibitem[\protect\citename{Wills, Netzer \& Wills\ }1985]{WNW}
   Wills B. J., Netzer H., Wills D., 1985, ApJ, 288, 94
\end{thebibliography}
\end{document}